# A Quantum-based Database Query Scheme for Privacy Preservation in Cloud Environment

Wenjie Liu[1, 2 *], Peipei Gao[2], Zhihao Liu[3, 4], Hanwu Chen[3, 4], Maojun Zhang[5, 6]

[1]Jiangsu Engineering Centre of Network Monitoring, Nanjing University of Information Science and Technology, Nanjing, China
[2]School of Computer and Software, Nanjing University of Information Science and Technology, Nanjing, China
[3] School of Computer Science and Engineering, Southeast University, Nanjing, China
[4] Key Laboratory of Computer Network and Information Integration (Southeast University), Ministry of Education，Nanjing, China
[5]School of Mathematics and Computing Science, Guilin University of Electronic Technology, Guilin, China
[6] College of Computer Science and Information Technology, Guangxi Normal University, Guilin, China
[*]Corresponding author: wenjiel@163.com

**Abstract:** Cloud computing is a powerful and popular information technology paradigm that enables data service outsourcing and provides higher-level services with minimal management effort. However, it is still a key challenge to protect data privacy when a user accesses the sensitive cloud data. Privacy-preserving database query allows the user to retrieve a data item from the cloud database without revealing the information of the queried data item, meanwhile limiting user's ability to access other ones. In this study, in order to achieve the privacy preservation and reduce the communication complexity, a quantum-based database query scheme for privacy preservation in cloud environment is developed. Specifically, all the data items of the database are firstly encrypted by different keys for protecting server's privacy, and in order to guarantee the clients' privacy, the server is required to transmit all these encrypted data items to the client with the oblivious transfer strategy. Besides, two oracle operations, a modified Grover iteration, and a special offset encryption mechanism are combined together to ensure that the client can correctly query the desirable data item. Finally, performance evaluation is conducted to validate the correctness, privacy, and efficiency of our proposed scheme.

**Keywords:** Cloud database, quantum-based database query, privacy preservation, oblivious transfer, oracle operation, Grover iteration, offset encryption

## 1. Introduction

Cloud computing is a powerful computing paradigm that enables ubiquitous access to shared infrastructure resources and higher-level services. It has shown the remarkable advantage in load balancing, data access control, and resources sharing, for database management [1]. Benefiting from the cloud paradigm, an increasing number of individuals and groups choose to put their massive data (including private part) into the cloud.

In recent years, database outsourcing has become an important component of cloud computing [2], where data owners outsource data management to a service provider (i.e., cloud database), and this mode is also called Database-as-a-Service (DaaS) [3]. Cloud database provides users with capabilities to store and process their data in the cloud, which has the advantages of scalability and high availability that users can access data anytime, anywhere and anyway. However, all the data of data owner is stored in the cloud environment, and some sensitive data (e.g., health records, financial transactions, personal information) is at risk of being compromised. So, security and privacy have become the major challenges which inhibit the cloud computing wide acceptance in practice [4].

The privacy preservation is the main concern of cloud application, such as service recommendation [5-7], service quality prediction [8, 9], database query [10-16] etc. As an important research branch, the privacy-preserving database query (PPDQ) aims to protect database security and clients' privacy, while ensuring the correctness of database query. To be specific, any user can query data items from the cloud database without revealing its information, but his/her access to other data items is not permitted. There are a variety of techniques or methods for guaranteeing the privacy preservation of database query, such as homomorphic encryption (HE) [10, 11], attribute-based encryption (ABE) [12, 13, 14], and searchable encryption (SE) [15, 16], etc. Searchable encryption is a cryptographic system which offers secure search functions over encrypted data, which is considered to be a more effective technique to solve the problem of PPDQ. In 2000, Song et al. [15] proposed the first searchable encryption scheme based on symmetric key cryptography (SKC). Since then, other various SE schemes have been continuously proposed, such as public key cryptography (PKC)-based searchable encryption [17], secure ranked search over encrypted cloud data [18], and so on.

As we all know, the security of classic cryptography protocols, including most private query schemes (also named privacy-preserving database query schemes), are based on mathematical complexity, and its security is based on the fact that computing power is limited. However, with the prevalence of new distributed computing models (especially cloud computing), a normal user is given the super computing power far beyond a single computer. Therefore, these cryptography protocols based on computational complexity are facing serious challenges.

On the other hand, quantum computing demonstrates the superior parallel computing power that the classical paradigm can't match. For instance, Shor's algorithm [19] solves the problem of integer factorization in polynomial time, and Grover's algorithm [20] has a quadratic speedup to the problem of conducting a search through some unstructured database. Therefore, most classic cryptography protocols, including PPDQ schemes, are very vulnerable to the powerful quantum computer. Fortunately, quantum mechanics also provides a security mechanism against quantum attacks, and it holds the potential unconditional security based on some physical properties, such as non-cloning theorem, uncertainty principle, quantum entanglement, etc. With the application of quantum mechanics in the field of information processing, some research findings have been proposed, including quantum key distribution [21, 22], quantum secret sharing [23, 24], quantum key agreement [25, 26], quantum direct communication [27, 28], quantum stegonagraphy [29], quantum teleportation and remote state preparation [30-32], quantum sealed-bid auction [33, 34], delegating quantum computation [35], and quantum machine learning [36, 37].

With the above observations, the security of classic database query schemes is facing the dual challenge of cloud computing and quantum computing, while quantum mechanics has been proven to be an effective method for solving such problem. In this study, in order to implement the privacy-preserving database query in cloud environment, we utilize some physical properties of quantum mechanics to design a quantum-based database query scheme for privacy preservation (QBDQ) in cloud environment, and conduct its performance evaluation to show our scheme is feasible, secure and efficient. To be specific, our main contributions include the three following aspects.

1) We present a systematic framework for privacy preservation cloud database query scheme in the cloud environment.
2) A feasible QBDQ is designed through oblivious transfer, the offset encryption mechanism, oracle operation, and the modified Grover iteration to achieve the privacy preservation for the cloud database query and reduce its communication complexity.
3) The performance evaluation is conducted to verify the performance of our proposed QBDQ scheme, such as correctness, security, and efficiency.

The rest of this paper is organized as follows. In Section 2, we introduce the basic knowledge of quantum computing, while the framework of the privacy-preserving database query in cloud environment is presented. In Section 3, the problem of privacy-preserving database query in cloud environment is defined, and then the proposed QBDQ is elaborated step by step. Section 4 conduct the performance evaluation from the aspects of correctness, security, and efficiency. After that, Section 5 summarizes the related work on cloud database queries, SE, and quantum private queries. Finally, the conclusion of the paper and the prospection for future work are presented in Section 6.

## 2. Preliminaries

In this section, the basic knowledge of quantum computing is introduced firstly. Then, we introduced the principle of oblivious transfer (OT). And finally, a cloud computing framework for privacy preservation is designed.

### 2.1 Quantum computing

**(1) Quantum bit**

The classic bit is the smallest unit in the classic computer, and its value is either 0 or 1. Unlike classical computers, the smallest unit of quantum computers is qubit (quantum bit), which is the quantum analog of the classic bit. A qubit is a unit vector in a two-dimensional complex Hilbert space, and its Dirac notation is represented as follows:

$$|\varphi\rangle = \alpha|0\rangle + \beta|1\rangle, \qquad (1)$$

where $\alpha$ and $\beta$ are the probability amplitudes of the state $|\varphi\rangle$, and $|\alpha|^2 + |\beta|^2 = 1$. Since the vectors $|0\rangle$ and $|1\rangle$ are basis states and can be represented as follows,

$$|0\rangle = \begin{pmatrix} 1 \\ 0 \end{pmatrix} \quad |1\rangle = \begin{pmatrix} 0 \\ 1 \end{pmatrix}, \quad (2)$$

the qubit $|\varphi\rangle$ can be expressed in vector form $|\varphi\rangle = \begin{pmatrix} \alpha \\ \beta \end{pmatrix}$. In addition, the single qubit can be extended to multiple qubits, for example, an n-qubit system can exist in any superposed basis states

$$|\varphi\rangle = \alpha_0|0\rangle + \alpha_1|1\rangle + \cdots \alpha_{2^n-1}|2^n-1\rangle. \quad (3)$$

Here, $\sum_{i=0}^{2^n-1}|\alpha_i|^2 = 1$. Quantum states $\{|0\rangle, |1\rangle, \ldots, |2^n-1\rangle\}$ form a complete orthonormal basis in Hilbert space.

**(2) Unitary operator**

In a closed quantum system, the evolution of the system is characterized by a series of unitary operators, that is,

$$|\varphi'\rangle = U|\varphi\rangle, \quad (4)$$

where $UU^\dagger = U^\dagger U = I$, and $U^\dagger$ is the transpose conjugate of $U$. Each unitary operator corresponds to a quantum gate. Similar to a logic gate in classical calculations, the quantum gate can be represented in matrix form, and the quantum gate over a qubit is represented by a $2 \times 2$ unitary matrix. For instance, Pauli-X, Pauli-Z, and the Hadamard gate $H$ are important quantum operators over one qubit described in Eq. (5)

$$X = \begin{pmatrix} 0 & 1 \\ 1 & 0 \end{pmatrix} \quad Z = \begin{pmatrix} 1 & 0 \\ 0 & -1 \end{pmatrix} \quad H = \frac{1}{\sqrt{2}}\begin{pmatrix} 1 & 1 \\ 1 & -1 \end{pmatrix}. \quad (5)$$

**(3) Quantum measurement**

The quantum state is in a superposition state, and it must be measured to collapse to a basis state to obtain a result. Assuming that the quantum state is $|\varphi\rangle = \frac{1}{\sqrt{N}}\sum_{i=0}^{N-1}\alpha_i|i\rangle$ before measurement operator, quantum measurements are described by a collection $\{M_i\}$ of measurement operators which satisfy the completeness equation

$$\sum_{i=0}^{N-1} M_i^\dagger M_i = I, \quad (6)$$

where $i$ indicates the possible outcome of the measurement. The quantum state is measured by the measurement basis $|i\rangle$, then the probability that result $i$ occurs is given by

$$p(i) = \langle\varphi|M_i^\dagger M_i|\varphi\rangle, \quad (7)$$

and the post-measurement state is

$$\frac{M_i|\varphi\rangle}{\sqrt{\langle\varphi|M_i^\dagger M_i|\varphi\rangle}}. \quad (8)$$

## 2.2 Oblivious transfer

In cryptography, an oblivious transfer (OT) strategy is a type of strategy in which a sender transfers one of potentially many pieces of information to a receiver, but remains oblivious as to what piece (if any) has been transferred. The first form of oblivious transfer was introduced by Rabin [38]. In this form, the sender sends a message to the receiver with probability 1/2, while the sender remains oblivious as to whether or not the receiver received the message. OT is a basic strategy in the field of cryptography and has a wide range of applications. In general, the OT strategy involves two parties, the sender and the receiver, and satisfies the following characteristics:

- Whether the queried data can be obtained is entirely dependent on probability, rather than sender or receiver. That is, neither the sender nor receiver can affect the execution of the strategy.
- After the execution of the strategy, the sender could not know whether the receiver got the data he wanted to query.

k-out-of-n ($OT_n^k$) (k<n) is the general form of all OT strategies. That is, the sender has n secrets, and the receiver can only get k secrets. The $OT_n^k$ strategy consists of two parties, the sender with n secret data $(d_0, d_1, \ldots, d_{n-1})$, and the receiver with k indices $(i_1, i_2, \ldots, i_k)$. The strategy meets the following requirements:

- **Correctness:** After executing the strategy, the receiver can obtain all of the $d_i$ correctly.
- **Receiver's security:** When the receiver queries the data from the sender, the database cannot know the receiver's query items.
- **Sender's security:** The receiver cannot get more data items from the sender except queried data items

## 2.3 The framework of privacy-preserving database query in cloud environment

We first consider the framework model of privacy-preserving cloud database query system, which consists of two main entities (clients and cloud server) as illustrated in Fig. 1.

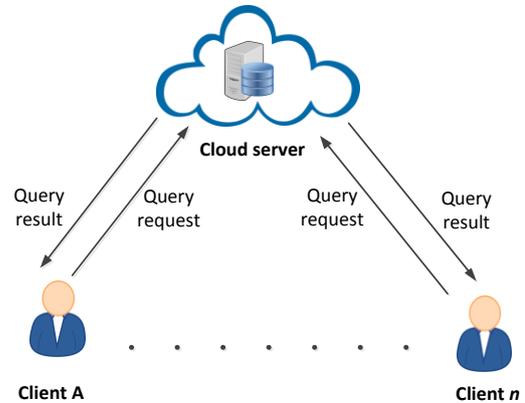

Fig. 1. The framework of privacy-preserving database query in cloud environment.

As shown in Fig. 1, there are $n$ clients and a cloud database server, and every client sends a query request to the cloud server and gets the query result from the cloud server finally. In this framework, we suppose all the clients and server are semi-honest: they are curious about cheating the privacy of other's, but honest to carry out the operations in the scheme. Here, two kinds of entity can be defined as below,

**Client** is the entity that wants to query items from the database in the cloud server and can be the connected users or the individual user with mobile constrained devices such as smartphones, PDA, TPM chip, etc.

**Cloud server** is the entity which provides data services and computational resources to the clients dynamically.

In this paper, we take three parties as an example, i.e., the client Alice, client Bob, and the cloud server Charlie, to demonstrate the process of the privacy-preserving database query using quantum mechanics.

## 3. A quantum-based database query scheme for privacy preservation in cloud environment

In this section, we first define the privacy-preserving database query problem and quantum-based privacy-preserving database query problem in cloud environment. To address this issue, a QBDQ scheme is proposed in detail. Before we introduce the relevant content, the key notations and descriptions used in this section are listed in Table 1.

Table 1. Key notations and descriptions involved in proposed QBDQ scheme

| Notation | Description |
|---|---|
| $N$ | The number of items in cloud server's database |
| $D$ | The cloud server's database, $D=\{D_0, D_1, ..., D_{N-1}\}$ |
| $D_i$ | The $i$-th data in $D$ |
| $n$ | The number of index qubits used to encode index of data items $n = \lceil \log N \rceil$ |
| $m$ | The number of qubits used to encode data $D_i$ |
| $p$ | The index of client Alice's query data |
| $D_p$ | The data item Alice wants to query from $D$ |
| $q$ | The index of client Bob's query data |
| $D_q$ | The data item Bob wants to query from $D$ |
| $\Delta s_A$ | The offset value of Alice |
| $\Delta s_B$ | The offset value of Bob |
| $K$ | The encryption key sequence belongs to Charlie, $K = \{K_0, K_1, ..., K_{N-1}\}$. |
| $O_K$ | The oracle operation to encode Charlie's key sequence $K$ |
| $O_D^A$ | The oracle operation to encode Alice's query result $D_p$ |
| $O_D^B$ | The oracle operation to encode Bob's query result $D_q$ |
| $O_s$ | The oracle operation which conditionally changes the sign in the amplitudes of the query item $D_p$ ($D_q$) |
| $O_p$ | The oracle operation which perform a conditional phase shift of -1 with every computational bass state except $|0\rangle$ |

### 3.1 Some definitions

In order to clearly illustrate our scheme, we first define the problem to be solved.

**Definition 1 (Database query problem for privacy preservation in cloud environment):** *In the cloud environment, the cloud server has a collection of sensitive data $D = \{D_0, D_1, ..., D_{N-1}\}$, and each client wants to query a data item $D_i$ ($0 \leq i \leq N-1$) from the cloud server without revealing which item is queried. During the retrieving process, the client cannot gain any other data item except $D_i$.*

**Definition 2 (Database query scheme for privacy preservation in cloud environment):** *Each client inputs the index of query item $i$ ($0 \leq i \leq N-1$), and cloud server inputs sensitive dataset $D = \{D_0, D_1, ..., D_{N-1}\}$. After executing this scheme, the client outputs the queried data item $D_i$. In addition, the scheme should satisfy:*

- ***Correctness:*** *The client successfully obtains the correct data item he(she) wants to query (i.e., $D_i$).*
- ***Clients' privacy:*** *During the retrieving process, the cloud server cannot get any private information about the query index of the client.*
- ***Cloud server's privacy:*** *Clients cannot get any other data items from the cloud server except $D_i$.*

### 3.2 A quantum-based database query scheme for privacy preservation in cloud environment

For the sake of simplicity, we take three parties (one cloud server Charlie, and two clients Alice, Bob) as an example to describe our scheme. Suppose Charile has a private database $D$ with $N$ items $\{D_0, D_1, ..., D_{N-1}\}$ and an encryption key sequence $K = \{K_0, K_1, ..., K_{N-1}\}$, and Alice and Bob want to respectively query an item, the $p$-th item $D_p$ and the $q$-th item $D_q$ ($0 \leq p, q \leq N-1$), from server. The scheme consists of five steps as follows (also shown in Fig. 2).

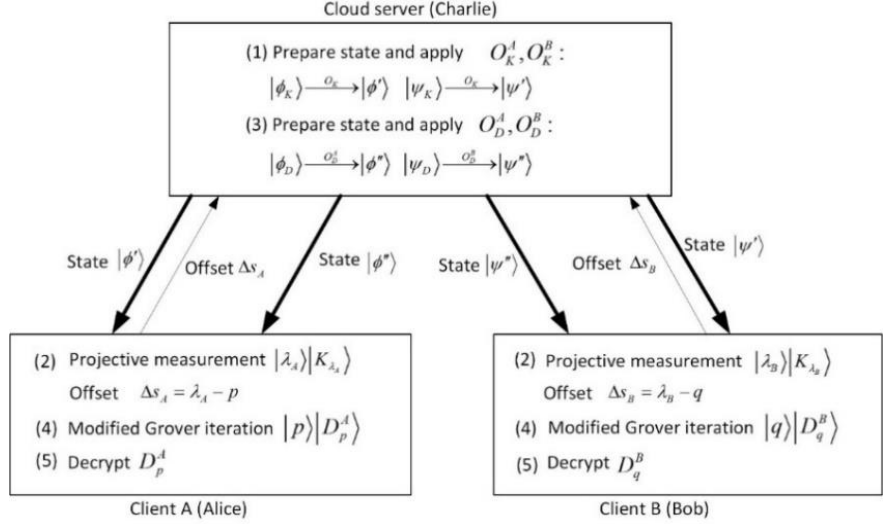

Fig. 2. The five-step procedures of the QBDQ scheme among two clients and cloud server. The thick (thin) line represents quantum (classic) channel.

**Step 1:** Charlie prepares an $(n+m)$-qubit state $|\phi_K\rangle = \frac{1}{\sqrt{N}}\sum_{i=0}^{N-1}|i\rangle\otimes|0\rangle^{\otimes m}$, where $n = \lceil logN \rceil$, $m = \lceil \log(\max\{k_i|0 \leq i \leq N-1\}+1)\rceil$. And then he applies an oracle operation $O_K$ (its schematic circuit is sketched in Fig. 3) on $|\phi_K\rangle$ referring to the sequence $K = \{K_0, K_1, \ldots, K_{N-1}\}$. Here, $O_K$ is defined as follows,

$$O_K : \frac{1}{\sqrt{N}}\sum_{i=0}^{N-1}|i\rangle\otimes|0\rangle^{\otimes m} \to \frac{1}{\sqrt{N}}\sum_{i=0}^{N-1}|i\rangle\otimes|K_i\rangle, \quad (9)$$

where $|i\rangle$ denotes the index of the data item, and $|K_i\rangle$ is the encryption key originally assigned to encrypt the $i$-th data item. After the above operation, we can get the state namely $|\phi'\rangle = \frac{1}{\sqrt{N}}\sum_{i=0}^{N-1}|i\rangle\otimes|K_i\rangle$, and then Charlie sends it to Alice with oblivious transfer strategy. Similar to Alice, Charlie also prepares another state $|\psi'\rangle = \frac{1}{\sqrt{N}}\sum_{i=0}^{N-1}|i\rangle\otimes|K_i\rangle$ in the same way and sends it to Bob.

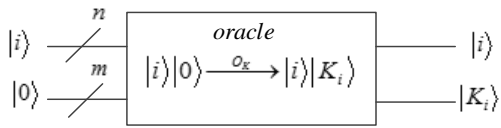

Fig. 3. Schematic circuit of the oracle operation $O_K$.

**Step 2:** After receiving $|\phi'\rangle$ from Charlie, Alice takes $\{|0\rangle, |1\rangle, \ldots, |N-1\rangle\}$ as the computational basis, and performs projective measurement on the index qubits of $|\phi'\rangle$. Suppose the measurement result is $\lambda_A$ ($\lambda_A \in \{0,1,\ldots,N-1\}$), the remaining $m$ qubits will collapse into $|K_{\lambda_A}\rangle$, which means Alice can obtain $K_{\lambda_A}$ (i.e., one of the encryption keys) through projective measurement. Since Alice's retrieving index is $p$, she computes the offset $\Delta s_A = (\lambda_A - p)$, and sends it to Charlie. As same as Alice, Bob also performs the same operations and announces the offset $\Delta s_B = (\lambda_B - q)$ to Charlie, where $\lambda_B$ is the measurement result, and $q$ represents the index of the data item Bob wants to query.

**Step 3:** Having received the offsets $\Delta s_A$ and $\Delta s_B$, Charlie updates every encryption key as follows,

$$\begin{aligned} K_i^A &= K_{(i+\Delta s_A)\bmod N} \\ K_i^B &= K_{(i+\Delta s_B)\bmod N} \end{aligned}, \quad (10)$$

and obtains the new key sequence $K^A$ and $K^B$,

$$\begin{aligned} K^A &= \{K_i^A \mid 0 \leq i \leq N-1\} \\ K^B &= \{K_i^B \mid 0 \leq i \leq N-1\} \end{aligned}. \quad (11)$$

Then, Charlie encrypts every data items respectively with its new corresponding keys $K_i^A$ and $K_i^B$ as below,

$$\begin{aligned} D_i^A &= D_i \oplus K_i^A, 0 \leq i \leq N-1 \\ D_i^B &= D_i \oplus K_i^B, 0 \leq i \leq N-1 \end{aligned}. \quad (12)$$

After that, Charlie prepares two states $|\phi_D\rangle = \frac{1}{\sqrt{N}}\sum_{i=0}^{N-1}|i\rangle\otimes|0\rangle^{\otimes m}$, $|\psi_D\rangle = \frac{1}{\sqrt{N}}\sum_{i=0}^{N-1}|i\rangle\otimes|0\rangle^{\otimes m}$, and applies the oracle operation $O_D^A$, $O_D^B$ as below,

$$\begin{aligned} O_D^A &: \frac{1}{\sqrt{N}}\sum_{i=0}^{N-1}|i\rangle|0\rangle^{\otimes m} \to \frac{1}{\sqrt{N}}\sum_{i=0}^{N-1}|i\rangle|D_i^A\rangle \\ O_D^B &: \frac{1}{\sqrt{N}}\sum_{i=0}^{N-1}|i\rangle|0\rangle^{\otimes m} \to \frac{1}{\sqrt{N}}\sum_{i=0}^{N-1}|i\rangle|D_i^B\rangle \end{aligned}, \quad (13)$$

and gets the final states $|\phi''\rangle = \frac{1}{\sqrt{N}}\sum_{i=0}^{N-1}|i\rangle|D_i^A\rangle$, $|\psi''\rangle = \frac{1}{\sqrt{N}}\sum_{i=0}^{N-1}|i\rangle|D_i^B\rangle$. Finally, Charlie sends $|\phi''\rangle$, $|\psi''\rangle$ to Alice and Bob, respectively with oblivious transfer strategy.

**Step 4:** After receiving $|\phi''\rangle$ from Charlie, Alice performs the modified Grover iteration on it to obtain the target state $|p\rangle|D_p^A\rangle$. Fig. 4 describes the detailed process of modified Grover iteration, which consists of at most $\left\lceil \frac{\pi}{4}\sqrt{2^{n+m}} \right\rceil$ times application of a quantum subroutine, called the $G$ operator. The whole process of $G$ operator (also shown in Fig. 5) can be subdivided into four steps as follows.

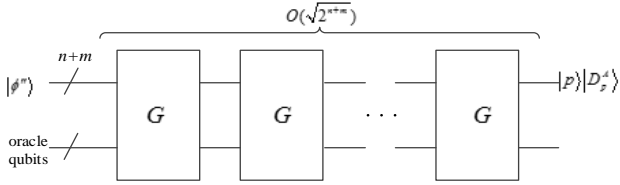

Fig. 4. Schematic circuit of the modified Grover iteration applied on state $|\phi''\rangle$, where $G$ is the quantum subroutine illustrated in Fig. 5.

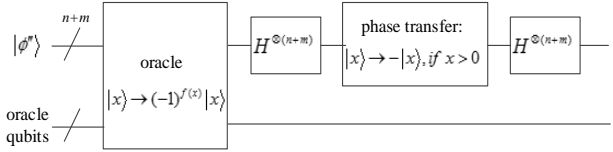

Fig. 5 Schematic circuit of the $G$ operator.

**Step 4-1:** Alice applies the oracle operation $O_s$ on $|\phi''\rangle$, which conditionally changes the sign of the amplitudes of the query item

$$O_s: \frac{1}{\sqrt{N}}\sum_{i=0}^{N-1}|i\rangle|D_i^A\rangle \rightarrow \frac{1}{\sqrt{N}}\sum_{i=0}^{N-1}(-1)^{f(i)}|i\rangle|D_i^A\rangle \quad (14)$$

Here, we call the resultant state $O_s|\phi''\rangle$, i.e., $O_s|\phi''\rangle = \frac{1}{\sqrt{N}}\sum_{i=0}^{N-1}(-1)^{f(i)}|i\rangle|D_i^A\rangle$, and $f(i)$ is the judgement function defined by:

$$f(i) = \begin{cases} 1, & \text{if } i \text{ is the query address } (i.e., i = p) \\ 0, & \text{else } (i.e., i \neq p) \end{cases} \quad (15)$$

**Step 4-2:** The Hadamard transformation $H^{\otimes(n+m)}$ is applied on $O_s|\phi''\rangle$,

$$\frac{1}{\sqrt{N}}\sum_{i=0}^{N-1}(-1)^{f(i)}|i\rangle|D_i^A\rangle \\ \rightarrow H^{\otimes(n+m)}\frac{1}{\sqrt{N}}\sum_{i=0}^{N-1}(-1)^{f(i)}|i\rangle|D_i^A\rangle \quad (16)$$

**Step 4-3:** Alice applies conditionally phase transfer $O_p$ on the state $H^{\otimes(n+m)}O_s|\phi''\rangle$,

$$O_p: H^{\otimes(n+m)}\frac{1}{\sqrt{N}}\sum_{i=0}^{N-1}(-1)^{f(i)}|i\rangle|D_i^A\rangle \\ \rightarrow -(-1)^{\sigma_{i,D_i}}H^{\otimes(n+m)}\frac{1}{\sqrt{N}}\sum_{i=0}^{N-1}(-1)^{f(i)}|i\rangle|D_i^A\rangle, \quad (17)$$

where the function $\sigma_{i,D_i}$ is defined as follows,

$$\sigma_{i,D_i} = \begin{cases} 1, & i = 0, D_i = 0, \\ 0, & \text{else.} \end{cases} \quad (18)$$

**Step 4-4:** The Hadamard transformation $H^{\otimes(n+m)}$ is applied again on $O_pH^{\otimes(n+m)}O_s|\phi''\rangle$, and obtains the state

$$|\xi\rangle = H^{\otimes(n+m)}O_PH^{\otimes(n+m)}O_s\frac{1}{\sqrt{N}}\sum_{i=0}^{N-1}|i\rangle|D_i^A\rangle \quad (19)$$

Alice applies the above Grover iteration $\left\lceil \frac{\pi}{4}\sqrt{2^{n+m}} \right\rceil$ times, and finally obtains the target state $|p\rangle|D_p^A\rangle$.

Similar to Alice, Bob also applies the modified Grover iteration on the received state $|\psi''\rangle = \frac{1}{\sqrt{N}}\sum_{i=0}^{N-1}|i\rangle|D_i^A\rangle$, and obtains the target query state $|q\rangle|D_q^B\rangle$.

**Step 5:** Alice and Bob measure the last $m$-qubit of state $|p\rangle|D_p^A\rangle$, $|q\rangle|D_q^B\rangle$, and extract the classic information of query result $D_p^A$, $D_q^B$, respectively.

In addition, in order to check eavesdropping in the quantum channel, we can use decoy-photon technology. That is, the sender randomly inserts several decoy photons into the qubit sequence, where every decoy photon is prepared randomly with either $Z$-basis $\{|0\rangle, |1\rangle\}$ or $X$-basis $\left\{\frac{1}{\sqrt{2}}(|0\rangle + |1\rangle), \frac{1}{\sqrt{2}}(|0\rangle - |1\rangle)\right\}$, and transmits them to the receiver. After confirming that the receiver has received the transmitted sequence, the sender announces the positions of the decoy photons and the corresponding measurement basis. The receiver measures the decoy photons according to the sender's announcements, and tells the sender his (her) measurement results. Then, the sender compares the measurement results from the receiver with the initial states of the decoy photons in the transmitted sequence, and calculates the error rate. If the error rate is higher than the threshold which determined by the channel noise, they cancel this scheme and restart; else they continue the next step.

It is worth mentioning that, we adopted the OT strategy and offset encryption mechanism in our scheme. In step 3, the OT strategy is utilized to transfer Charlie's data to Alice and Bob. As we know, the transmitted state $|\psi\rangle$ is a superposition state which encapsulates all the encrypted data items $\{D_i | 0 \leq i \leq N-1\}$. So, the process of Charlie sending $|\phi'\rangle$, $|\psi'\rangle$ to Alice and Bob can be viewed as the oblivious transfer mechanism. The use of OT strategy ensures that information about Charlie cannot be leaked. In addition, our scheme also applied the offset encryption mechanism. The offsets $\Delta s_A$, $\Delta s_B$ can be computed by using the index of the query data items and the keys determined by clients' measurement. Charlie updates the encryption keys according to these offsets, and then encrypts data with these updated keys, respectively. The combination of OT strategy and offset mechanism allows Alice and Bob obtain the correct data they want to query, while Charlie cannot get their queried data, which guaranteed the privacy of client. At the same time, data encryption makes the data items into ciphertext, and neither the eavesdropper nor the clients can directly obtain the data item, thus ensuring the data security of the cloud server.

## 4. Performance evaluation

Our proposed QBDQ scheme in cloud environment tends to ensure the correctness of query result, protect the privacy of clients and servers in cloud, and also improve the efficiency during querying the cloud database. Therefore, we take three parties (i.e., clients Alice and Bob, cloud server Charlie) as an example, and estimate the overall performance of the proposed scheme in terms of correctness analysis, security analysis, and the efficiency analysis.

### 4.1 Correctness analysis

Now, we analyze the correctness of the proposed scheme. Without loss of generality, suppose that the server Charlie has a database of 16 items $D = \{5,9,6,12,2,11,11,6,5,10,7,15,6,11,6,9\}$, and he holds the corresponding encryption key sequence $K = \{14,8,3,4,7,1,11,6,15,2,12,13,0,5,9,10\}$. Since $N = 16$, the max value in $K$ is 15, $n = \lceil \log N \rceil = \lceil \log 16 \rceil = 4$, $m = \lceil \log(15+1) \rceil = 4$. Here, we take Alice as an example to analyze the procedures of our QBDQ scheme as below (suppose Alice want to query the 9th item of the database).

In step 1, Charlie prepares an initial state $|\phi_K\rangle = \frac{1}{4}\sum_{i=0}^{15}|i\rangle \otimes |0000\rangle$ and performs an oracle operation $O_K$ on it to encode his encryption keys,

$$|\phi'\rangle = \frac{1}{4}(|0\rangle|K_0\rangle + |1\rangle|K_1\rangle + |2\rangle|K_2\rangle + |3\rangle|K_3\rangle + |4\rangle|K_4\rangle + |5\rangle|K_5\rangle$$
$$+|6\rangle|K_6\rangle + |7\rangle|K_7\rangle + |8\rangle|K_8\rangle + |9\rangle|K_9\rangle + |10\rangle|K_{10}\rangle + |11\rangle|K_{11}\rangle$$
$$+|12\rangle|K_{12}\rangle + |13\rangle|K_{13}\rangle + |14\rangle|K_{10}\rangle + |15\rangle|K_{11}\rangle)$$
$$= \frac{1}{4}(|0\rangle|14\rangle + |1\rangle|8\rangle + |2\rangle|3\rangle + |3\rangle|4\rangle + |4\rangle|7\rangle + |5\rangle|1\rangle + |6\rangle|11\rangle$$
$$+|7\rangle|6\rangle + |8\rangle|15\rangle + |9\rangle|2\rangle + |10\rangle|12\rangle + |11\rangle|13\rangle + |12\rangle|0\rangle$$
$$+|13\rangle|5\rangle + |14\rangle|9\rangle + |15\rangle|10\rangle)$$

(20)

Then, he sends the resultant state $|\phi'\rangle$ to Alice. In step 2, Alice performs projective measurement on the first four qubits (i.e., index qubits) of $|\phi'\rangle$ in the computational basis $\{|0000\rangle, |0001\rangle, \ldots, |1111\rangle\}$. Suppose the random measurement result is $|12\rangle$ (i.e., $\lambda_A = 12$), then the remaining qubits (i.e., the key qubits) collapse to the state $|K_{\lambda_A}\rangle = |0000\rangle$, which means $K_{\lambda_A} = 0000$. But the data Alice wants to query is the ninth data $D_8$, so she computes the difference between $\lambda_A$ and the desirable query index q, $\Delta s = (\lambda_A - q) = 4$, and sends $\Delta s$ to Charlie. After receiving $\Delta s$, Charlie updates the key sequence K through the formulation $K_i^A = K_{(i+\Delta s) \bmod N}$, then $K_i^A = \{7,1,11,6,15,2,12,13,0,5,9,10,14,8,3,4\}$. He uses $K_i^A$ to encrypt every data items: $D_i^A = D_i \oplus K_i^A$, that is, $\{2,8,13,10,13,9,7,11,5,15,14,5,8,3,5,13\}$. Then, in step 3, Charlie prepares another state $|\phi_D\rangle = \frac{1}{4}\sum_{i=0}^{15}|i\rangle \otimes |0000\rangle$ and applies the oracle operation $O_D^A$ to embed the encrypted data items $D_i^A$,

$$|\phi''\rangle = \frac{1}{4}(|0\rangle|D_0\rangle + |1\rangle|D_1\rangle + |2\rangle|D_2\rangle + |3\rangle|D_3\rangle + |4\rangle|D_4\rangle + |5\rangle|D_5\rangle$$
$$+|6\rangle|D_6\rangle + |7\rangle|D_7\rangle + |8\rangle|D_8\rangle + |9\rangle|D_9\rangle + |10\rangle|D_{10}\rangle + |11\rangle|D_{11}\rangle$$
$$+|12\rangle|D_{12}\rangle + |13\rangle|D_{13}\rangle + |14\rangle|D_{14}\rangle + |15\rangle|D_{15}\rangle)$$
$$= \frac{1}{4}(|0\rangle|2\rangle + |1\rangle|8\rangle + |2\rangle|13\rangle + |3\rangle|10\rangle + |4\rangle|13\rangle$$
$$+|5\rangle|9\rangle + |6\rangle|7\rangle + |7\rangle|11\rangle + |8\rangle|5\rangle + |9\rangle|15\rangle$$
$$+|10\rangle|14\rangle + |11\rangle|5\rangle + |12\rangle|8\rangle + |13\rangle|3\rangle$$
$$+|14\rangle|5\rangle + |15\rangle|13\rangle)$$

(21)

Then, he sends the state $|\phi''\rangle$ to Alice.

Further, Alice performs modified Grover iteration on $|\phi''\rangle$ up to $R = \lceil \frac{\pi}{4}\sqrt{2^{n+m}} \rceil = \lceil \frac{\pi}{4}\sqrt{2^8} \rceil = 13$ times (actually, the number of iterations is 6), then she can obtain the encrypted query item $|p\rangle|D_p^A\rangle = |8\rangle|5\rangle$ with high possibility, and measures it to get $D_p^A = 5$. Alice uses the obtained key $K_8^A = 0$ to decrypt the ninth item

$$D_8 = D_8^A \oplus K_8^A = 0101 \oplus 0000 = 0101. \quad (22)$$

Therefore, regardless of what measurement result Alice has obtained, she can finally obtain the query data correctly.

Fig. 6 shows the entire execution process of Alice querying Charlie's database in a simplified way. At the same time, it also sketched the execution of the other user Bob (assuming it queries the fifth data).

### 4.2 Security analysis

**(1) Privacy analysis**

**Cloud Server's privacy.** Suppose the client Alice is dishonest, and she wants to obtain more information about Charlie's database. In step 1 of our scheme, the server Charlie sends the quantum state $|\phi'\rangle = \frac{1}{\sqrt{N}}\sum_{i=0}^{N-1}|i\rangle|K_i\rangle$ to client Alice through oblivious transfer strategy. Since all the information about the key sequence $K$ is encoded in the state $|\phi'\rangle$, so Alice cannot extract the key form $|\phi'\rangle$ directly. Here we suppose the whole system of quantum state $|\phi'\rangle$ consisted of two subsystems, i.e., the $n$-qubit quantum subsystem $C$ (index qubits $|i\rangle$) and the $m$-qubit subsystem $D$ (key qubits $|K_i\rangle$). If Alice makes a projective measurement on the received state $|\phi'\rangle$, she will get the resultant state $|i\rangle|K_i\rangle$ for any $i$ with the probability of $\frac{1}{N}$. The whole system can be represented by the quantum ensemble

$\varepsilon = \{p_i, \rho(i)\}$, here $p_i = \frac{1}{N}$,
$$\rho(i) = |i\rangle |K_i\rangle \langle K_i| \langle i|. \quad (23)$$
Here we get the upper limit of information that Alice can get from Charlie's is determined by the Holevo bound [38],
$$H(A:B) \leq S(\rho) - \frac{1}{N} \sum_{i=0}^{N-1} S(\rho(i)) \quad (24)$$
Here $S(\rho)$ denotes Von Neumann entropy of quantum state $\rho$, $H(B:A)$ means the information Alice can get about Charlie's key information (including the address $i$ and according keys $K_i$), we have
$$S(\rho) = S(\frac{1}{N} \sum_{i=0}^{N-1} |i\rangle |K_i\rangle \langle K_i| \langle i|) = n + m. \quad (25)$$
and $S(\rho(i)) = S(|i\rangle |K_i\rangle \langle K_i| \langle i|) = 0$, therefore,
$$H(A:B) \leq n + m. \quad (26)$$
Then, Alice can only get $n$-bit of address information (i.e., $i$) and the corresponding $m$-bit key(i.e., $K_i$) by measuring $\rho$. In addition, she will certainly lose the change to get her key $K_i$. This means Alice cannot extract more than one key from Charlie.

Besides, in step 3, Charlie uses the offset key $K_i^A = K_{i+\Delta s}$ to encrypt the data items, and send its encoded state $|\phi''\rangle = \frac{1}{\sqrt{\Delta s}} \sum_{i=0}^{N-1} |i\rangle |D_i^A\rangle$ to Alice with oblivious transfer strategy. Alice's privacy of query index $i$ is protected by the oblivious transfer strategy. For example, the transmitted state $|\phi''\rangle$ Alice received is a superposition state, i.e., $|\phi''\rangle = \frac{1}{\sqrt{N}}(|0\rangle|D_0\rangle + |1\rangle|D_1\rangle + \cdots + |N-1\rangle|D_{N-1}\rangle)$, which encapsulates all the query data $\{D_p^A | 0 \leq p \leq N-1\}$ including the desirable one $D_p^A$. Alice obtains the query item $D_p^A$ through the Grover iteration and the previously obtained key $K_{\lambda_A}$. Suppose Bob is also dishonest, he has the same situation with Alice.

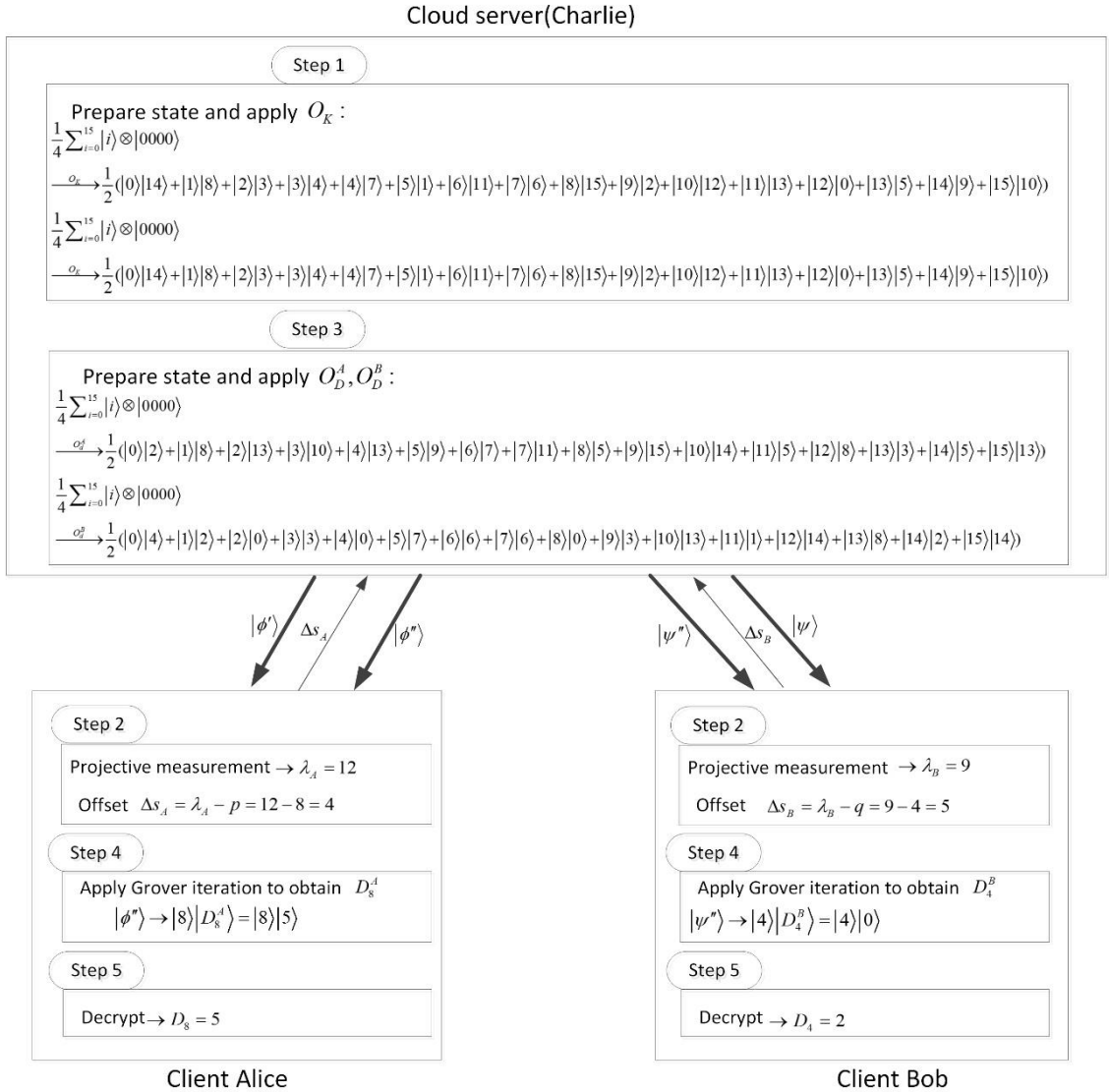

Fig. 6 The schematic graph of the execution process of Alice and Bob in our QBDQ scheme, assuring they query the *9-th* and *5-th* items, respectively.

**Client's privacy.** If Charlie is dishonest, he may try to obtain Alice's private query index $p$ during the communication process. However, Alice only sends one classic message $\Delta s = \lambda_A - p$ to cloud server Charlie in Step 2, and Charlie does not know the encryption key which chosen by Alice, thus he cannot obtain any useful information about the data Alice wants to search. As same as Alice, Bob only sends a classic offset message $\Delta s = \lambda_B - p$ to Charlie, which prevents Charlie from obtaining his information.

**(2) Channel security analysis**

The security of the quantum channel is guaranteed by the decoy-photon checking technology. The process of eavesdropping detect done by the two neighbor participants in our scheme is essentially equivalent to that in the BB84 scheme [36], which has been proved to be unconditionally secure. To be specific, the decoy qubits, which are randomly inserted into target qubits, are generated by randomly chosen from $\{|0\rangle, |1\rangle, |+\rangle, |-\rangle\}$. After one participant sends the mixed decoy qubits and encrypted target qubits to quantum center, he will ask quantum center to measure them with the same bases these qubits were produced. For any outside eavesdropper, the bases used by participants are all random, the eavesdropper cannot produce the same qubits like decoy qubits before quantum center receives the qubits. Just like the situation in the BB84 scheme, if any outside eavesdropper exists in the process of our scheme, the eavesdropping actions will be found by the two participants.

The outside eavesdropper cannot get the shared key because eavesdropper cannot distinguish target qubits form decoy qubits, and he can only choose one set of orthogonal basis to measure it, so the eavesdropper will certainly change the states of the qubit, and then he will be discovered. We assume that eavesdropper will do intercept-resend attack. Eavesdropper applies operation $U_E$ and auxiliary system $|E\rangle$ which satisfies the following conditions,

$$U_E|0\rangle|E\rangle = a|0\rangle|E_{00}\rangle + b|1\rangle|E_{01}\rangle, \quad (27)$$

$$U_E|1\rangle|E\rangle = c|0\rangle|E_{01}\rangle + d|1\rangle|E_{11}\rangle. \quad (28)$$

Here, $|a|^2 + |b|^2 = 1$ and $|c|^2 + |d|^2 = 1$. If the eavesdropper wants to extract the encode information precisely, then $U_E$ must satisfy

$$U_E|+\rangle|E\rangle = \frac{1}{\sqrt{2}}(a|0\rangle|E_{00}\rangle + b|1\rangle|E_{01}\rangle + c|0\rangle|E_{10}\rangle + d|1\rangle|E_{11}\rangle)$$
$$= \frac{1}{2}(|+\rangle(a|E_{00}\rangle + b|E_{01}\rangle + c|E_{10}\rangle + d|E_{11}\rangle)), \quad (29)$$

$$U_E|-\rangle|E\rangle = \frac{1}{\sqrt{2}}(a|0\rangle|E_{00}\rangle + b|1\rangle|E_{01}\rangle - c|0\rangle|E_{10}\rangle - d|1\rangle|E_{11}\rangle)$$
$$= \frac{1}{2}(|-\rangle(a|E_{00}\rangle - b|E_{01}\rangle - c|E_{10}\rangle + d|E_{11}\rangle)), \quad (30)$$

$$U_E|+y\rangle|E\rangle = \frac{1}{\sqrt{2}}(a|0\rangle|E_{00}\rangle + b|1\rangle|E_{01}\rangle + ic|0\rangle|E_{10}\rangle + id|1\rangle|E_{11}\rangle)$$
$$= \frac{1}{2}(|+y\rangle(a|E_{00}\rangle - ib|E_{01}\rangle + ic|E_{10}\rangle - d|E_{11}\rangle)), \quad (31)$$

$$U_E|-y\rangle|E\rangle = \frac{1}{\sqrt{2}}(a|0\rangle|E_{00}\rangle + b|1\rangle|E_{01}\rangle - ic|0\rangle|E_{10}\rangle - id|1\rangle|E_{11}\rangle)$$
$$= \frac{1}{2}(|+y\rangle(a|E_{00}\rangle + ib|E_{01}\rangle - ic|E_{10}\rangle + d|E_{11}\rangle)). \quad (32)$$

From Eq. (29)-(32) we can obtain that

$$a|E_{00}\rangle - b|E_{01}\rangle + c|E_{10}\rangle - d|E_{11}\rangle = 0, \quad (33)$$

$$a|E_{00}\rangle + b|E_{01}\rangle - c|E_{10}\rangle - d|E_{11}\rangle = 0, \quad (34)$$

$$a|E_{00}\rangle + ib|E_{01}\rangle + ic|E_{10}\rangle - d|E_{11}\rangle = 0, \quad (35)$$

$$a|E_{00}\rangle - ib|E_{01}\rangle - ic|E_{10}\rangle - d|E_{11}\rangle = 0, \quad (36)$$

we can get that $a = d = 1$, $b = c = 0$ and $|E_{00}\rangle = |E_{11}\rangle$, then we get

$$U_E|0\rangle|E\rangle = |0\rangle|E_{00}\rangle, \quad (37)$$

$$U_E|1\rangle|E\rangle = |1\rangle|E_{11}\rangle, \quad (38)$$

we can summarize that eavesdropper would not be found only when decoy qubits and target qubits are $\{|0\rangle, |1\rangle\}$, which is impossible. So there is no way for the eavesdropper to know the secret key.

**4.3 Efficiency analysis**

As we know, Quantum-based schemes have greater information capacity than classic ones. In order to evaluate the efficiency of our QDBQ scheme more objectively, We choose some of the most representative quantum schemes as comparison objects, for example, Jakobi et al.'s quantum private query (QPQ) scheme (J11 for short) [44], Gao et al.'s QPQ scheme (G12) [45], and Rao et al.'s QPQ) scheme (R13) [46].

To evaluate the efficiency of quantum communication schemes, there are mainly two indicators: the communication complexity (i.e., the number of transmitted qubits), and the consumption of exchanged classic messages (i.e., the number of exchanged classic bits).

**(1) Communication complexity**

The communication complexity, i.e., the number of quantum bits (qubits) transmitted in the communication process, is one of the key indicators of the efficiency for communication scheme. In J11 and G12 schemes, the cloud server (Charlie) sends $k \times N$ qubits to the client (Alice), where $k$ is the number of divided substrings. These $k$ substrings are added bitwise in order to reduce Alice's information on the key to roughly one bit (i.e., $\bar{n} = N(\frac{1}{4})^k \approx 1$), so $k = \log\sqrt{N}$. In summary, $N \times \log\sqrt{N}$ qubits are transmitted in J11 and J12 schemes, and its communication complexity is $O(N\log N)$. But in the R13 scheme, the number of qubits that need to be exchanged is reduced to $O(N)$, so the communication complexity is $O(N)$.

In our QBDQ scheme, Charlie firstly transmits a $(\lceil logN \rceil + m)$-qubit state $|\phi'\rangle$ ($|\psi'\rangle$) for sending the encryption keys in step 1, and the $(\lceil logN \rceil + m)$-qubit state $|\phi''\rangle$ ($|\psi''\rangle$) containing every encrypted data $D_i(0 \leq i \leq N-1)$ is transmitted to Alice(Bob) in step 3. Considering that each data item the cloud server holds is an only one-bit message in J11, G12 and R13 schemes, here we let $m=1$. Therefore, the transmitted qubits are $2(logN + 1)$, so its communication complexity is $O(logN)$.

To be more intuitive, we calculate the numbers of transmitted qubits in different database capacities for the J11, G12, R13, and our QBDQ schemes (see Table 2), and show the comparison results among them in Fig. 7. As shown in this figure, J11 and G12 schemes have the same qubits consumption, R13 scheme reduces the consumption, and our QBDQ scheme has the lowest qubits consumption. That is, our scheme has the lowest communication complexity among them.

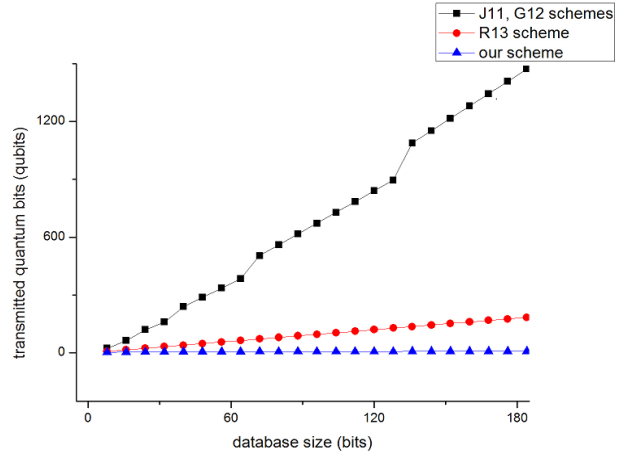

Fig. 7: Comparison of transmitted qubits among our QBDQ scheme, J11, G12, R13 schemes.

**(2) Consumption of exchanged classic messages**

For a communication scheme, it should also consider the consumption of the exchanged classic messages. In the J11, J12 and R13 schemes, $N \times 1$ bits of encrypted data, considering each data item is a one-bit message (i.e., $m = 1$), are transmitted from the cloud server to the client, so their exchanged classic messages are all $O(N)$ cbits. In our scheme, Alice (Bob) returns a classical message $\Delta s$, i.e., a $(\lceil logN \rceil \times m)$-cbit classic message, to Charlie in step 2. Since $m = 1$, the exchanged message is just $O(logN)$ cbits.

Table 3 lists the numbers of transmitted qubits in different database capacities for the J11, G12, R13, and our QBDQ schemes, while Fig. 8 gives a more intuitive comparison between our QBDQ scheme and the other QPQ schemes (J11, G12, R13 schemes). Obviously, our scheme needs less consumption of exchanged classic messages than other QPQ protocols.

Table. 2: Numbers of transmitted qubits in different database capacities for J11, G12, R13 and our QBDQ schemes.

| Database size | Transmitted messages(qubit) | | | Database size | Transmitted messages(qubit) | | |
|---|---|---|---|---|---|---|---|
| | J11/ G12 | R13 | QBDQ | | J11/G12 | R13 | QBDQ |
| 8 | 24 | 8 | 3 | 208 | 1664 | 208 | 8 |
| 16 | 64 | 16 | 4 | 216 | 1728 | 216 | 8 |
| 24 | 120 | 24 | 5 | 224 | 1792 | 224 | 8 |
| 32 | 160 | 32 | 5 | 232 | 1856 | 232 | 8 |
| 40 | 240 | 40 | 6 | 240 | 1920 | 240 | 8 |
| 48 | 288 | 48 | 6 | 248 | 1984 | 248 | 8 |
| 56 | 336 | 56 | 6 | 256 | 2048 | 256 | 8 |
| 64 | 384 | 64 | 6 | 264 | 2376 | 264 | 9 |
| 72 | 504 | 72 | 7 | 272 | 2448 | 272 | 9 |
| 80 | 560 | 80 | 7 | 280 | 2520 | 280 | 9 |
| 88 | 616 | 88 | 7 | 288 | 2592 | 288 | 9 |
| 96 | 672 | 96 | 7 | 296 | 2664 | 296 | 9 |
| 104 | 728 | 104 | 7 | 304 | 2736 | 304 | 9 |
| 112 | 784 | 112 | 7 | 312 | 2808 | 312 | 9 |
| 120 | 840 | 120 | 7 | 320 | 2880 | 320 | 9 |
| 128 | 896 | 128 | 7 | 328 | 2952 | 328 | 9 |
| 136 | 1088 | 136 | 8 | 336 | 3024 | 336 | 9 |
| 144 | 1152 | 144 | 8 | 344 | 3096 | 344 | 9 |
| 152 | 1216 | 152 | 8 | 352 | 3168 | 352 | 9 |
| 160 | 1280 | 160 | 8 | 360 | 3240 | 360 | 9 |
| 168 | 1344 | 168 | 8 | 368 | 3312 | 368 | 9 |
| 176 | 1408 | 176 | 8 | 376 | 3384 | 376 | 9 |
| 184 | 1472 | 184 | 8 | 384 | 3456 | 384 | 9 |
| 192 | 1536 | 192 | 8 | 392 | 3528 | 392 | 9 |
| 200 | 1600 | 200 | 8 | 400 | 3600 | 400 | 9 |

Table. 3: Exchanged classic messages in different database capacities for J11, G12, R13 and our QBDQ schemes.

| Database size | Exchanged messages | | Database size | Exchanged messages | |
|---|---|---|---|---|---|
| | J11/G12/R13 | QBDQ | | J11/G12/R13 | QBDQ |
| 8 | 8 | 3 | 88 | 88 | 7 |
| 16 | 16 | 4 | 96 | 96 | 7 |
| 24 | 24 | 5 | 104 | 104 | 7 |
| 32 | 32 | 5 | 112 | 112 | 7 |
| 40 | 40 | 6 | 120 | 120 | 7 |
| 48 | 48 | 6 | 128 | 128 | 7 |
| 56 | 56 | 6 | 136 | 136 | 8 |
| 64 | 64 | 6 | 144 | 144 | 8 |
| 72 | 72 | 7 | 152 | 152 | 8 |
| 80 | 80 | 7 | 160 | 160 | 8 |

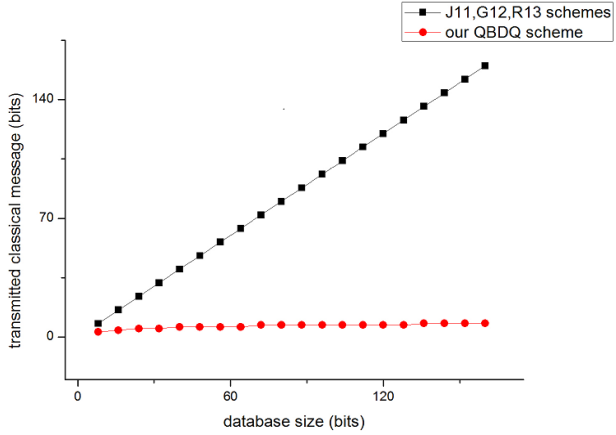

Fig. 8: Comparison of exchanged classic bits between our QBDQ scheme and the other QPQ schemes (J11, G12, R13 schemes).

In summary, Table 4 lists the comparison among our QBDQ scheme and the other three QPQ schemes clearly. As shown in Table 4, our scheme achieves a great reduction on both the communication complexity and the consumption of exchanged classic messages. Besides, our QBDQ scheme just needs to perform quantum measurement two times, which is obviously less than the other ones.

Table 4. Comparison among our QBDQ scheme and the other QPQ schemes

| Schemes | Communication complexity (qubit) | Exchanged message (bit) | Measurement times |
|---|---|---|---|
| J11 | $O(N\log N)$ | N+1 | kN |
| G12 | $O(N\log N)$ | N+1 | kN |
| R13 | $O(N)$ | N+1 | N |
| Ours | $O(\log N)$ | 1 | 2 |

## 5. Related Work

Cloud database services are typically run on cloud computing platforms, and access to cloud databases is provided as a service, which takes care of scalability and availability of the database, and it makes the underlying software-stack transparent to the user.

Benefit from cloud computing technologies and devices, more and more data owners are motivated to outsource their data to cloud servers for great convenience in data management, and cloud database query has attracted the attention of scholars. Cloud database query was firstly proposed by Chor et al. [40], where the privacy of the server cannot be guaranteed, which means that sensitive data (e.g., health records, financial transactions) stored in cloud database is threatened by information leaks. Therefore, how to preserve the privacy of sensitive data in the process of cloud database query has become an important topic. In order to solve the problem, many methods are proposed to guarantee the privacy preservation of database query [12-18], one of the most popular methods is SE.

SE is a special kind of private query, which enables the user to store the encrypted data to the cloud and execute keyword search over ciphertext. Since Song et al.[15] proposed the first practical private database query scheme for searching on encrypted data in cloud and provided the security proofs for the scheme, some other schemes to address privacy protection issues in cloud database queries have also been proposed[17, 18]. In order to support more complex queries, the conjunctive keyword search scheme [14] over encrypted data has been proposed. After that, a more general approach, predicate encryption [16], which supports inner-product, was also proposed.

In general, most of the above schemes [12-18, 40] are based on public key cryptography such as RSA, and its security is based on mathematical NP-hard problems. Therefore, these schemes are difficult to crack in polynomial time for a classic computers. all of the above protocols is based on public key cryptography such as RSA. On a quantum computer, to factor an integer $N$, Shor's algorithm [19] can run in polynomial time (the time taken is polynomial in $\log N$, which is the size of the input. Specifically, it takes quantum gates of order $O((\log N)^2(\log\log N)(\log\log\log N))$ using fast multiplication,[41] thus demonstrating that the integer-factorization problem (The large factorization problem is the security foundation of RSA) can be efficiently solved on a quantum computer and is consequently in the complexity class BQP. This is almost exponentially faster than the most efficient known classical factoring algorithm, so we can say that these schemes [12-18, 40] are not resistant to quantum attacks. Different from classic schemes based on mathematical complexity, the security of quantum based schemes is guaranteed by some properties of quantum mechanics, such as non-cloning theorem, uncertainty principle. They are considered to have potential unconditional security, and of course also include resistance to quantum attacks.

Recently, some researchers have tried to utilize quantum mechanics to design private query schemes. In 2008, Giovannetti et al. [42] proposed the first quantum private query (QPQ) protocol. The client sends the query $|j\rangle_Q$ and a decoy state $(|j\rangle_Q + |0\rangle_Q)/\sqrt{2}$ to the server in random order, then Bob uses each of them to interrogate his database using a qRAM (which records the reply to her queries in a register R), and returns $|j\rangle_Q|A_j\rangle_R$ or $(|j\rangle_Q|A_j\rangle_R + |0\rangle_Q|0\rangle_R)/\sqrt{2}$. The returned decoy state is used to check the eavesdropping of the server or the outside party. In 2011, Olejnik [43] presented a new QPQ protocol in a similar form with Giovannetti et al.'s protocol. By subtly selecting the oracle operation and the encoding scheme, one query state can achieve two aims simultaneously, i.e., obtaining the expected information and checking Bob's potential attack, so the communication complexity is reduced.

Unfortunately, it is very vulnerable to the realities of significant transmission losses.

Therefore, Jakobi et al. [44] proposed a novel QPQ protocol (J11) based on the QKD protocol, where QKD is essentially a quantum analog of SE. In this protocol, an asymmetric key can be distributed between Alice and Bob by utilizing SARG04 QKD protocol, and Bob encrypts the whole database with the QKD key. Alice only knows few bits of the key, which ensures the database privacy. Compared with the previous QPQ protocols, J11 protocol is loss-tolerant and more secure. What's more, the J11 protocol can be easily generalized to the large database. Later, Gao et al. [45] proposed a flexible generalized protocol (G12) based on the J11 protocol, which introduced a variable $\theta$ to adjust the balance between database security and client privacy. Considering a database with size $N$, the J11 and G12 protocols have a communication complexity of $O(N\log N)$. In order to reduce the complexity, Rao et al. [46] gave two more efficient protocols (R13), which reduced the number of exchanged qubits to $O(N)$.

Different from classical encryption schemes based on some mathematical difficult problems, these findings have shown the potential in either the improvements of efficiency or the enhancements of security in cloud computing field with large computing resources and also brought new quantum technologies to solve private database query problems. However, to the best of our knowledge, there are few studies focusing on the quantum-based privacy-preserving database query problem in cloud environment. Therefore, we combine quantum mechanics with cloud database queries and proposed a QBDQ which aims to realize the privacy preservation for the clients and cloud server.

## 6. Conclusion and Future Work

As far as we know, the existing QPQ schemes either belong to the qRAM-based schemes, such as, Giovannetti et al.'s [42] and Olejnik's [43] schemes, or belong to the QKD-based schemes, such as, Jakobi et al.'s [44], Gao et al.'s [45] and Rao et al.' s [46] schemes. These QKD-based schemes solve the problem of the server's privacy, their communication complexity needs to be further reduced. In this study, we propose an efficient quantum private query scheme based on oracle operation, modified Grover iteration, oblivious transfer strategy and the special offset encryption mechanism rather than QKD or qRAM. Compared with those schemes, our QBDQ scheme shows higher efficiency in terms of the communication complexity, the consumption of exchanged message, and the quantum measurement.

In our QBDQ scheme, we adopt the oblivious transfer strategy to solve the problem of the client's privacy, i.e., the client will ask the server to transmit all these encrypted data items to him/her. But in a real-world cloud environment, this is not a good approach. Although it guarantees that there is no information about the query index to be leaked, but it needs to transmit too many data items from the cloud database. Even if quantum resources have an exponential high-capacity advantage, it is also a waste of resources. Maybe the "query window" strategy is a better choice. To be specific, the client can firstly choose an index window that contains the desirable query item, and ask the server transmit these encrypted data items in this window scope other than the all data items, to him/her in a quantum way. Although there is certain information leakage from the perspective of information theory, it can save quantum resources. In this strategy, the selection of the size of a query window is a key point. In order to achieve a balance of efficiency and security, perhaps some game theory (such as, Nash Equilibrium [47, 48]) and penalty functions[49-51] can provide relevant optimized solutions.

It's worth noting that although the proposed solution involves two clients, for the sake of brevity (and for comparison with other quantum schemes), Alice and Bob do not interact. This is the most common pattern in cloud database queries. For the multi-party joint inquiry method, we will discuss it in future work. In addition, we just consider the ideal framework of the privacy-preserving database query in cloud environment, i.e., all the clients and cloud server are semi-honest. But in a real cloud environment, clients and servers may be untrustworthy. How to generalize our QBDQ into such multi-user and the untrusted scenario is an interesting work.

**Acknowledgement**

The authors would like to thank the anonymous reviewers and editor for their comments that improved the quality of this paper. This work is supported by Nature Science Foundation of China (Grant Nos. 71461005, 61502101, 61501247 and 61672290), Natural Science Foundation of Jiangsu Province (Grant Nos. BK20171458), Natural Science Foundation for Colleges and Universities of Jiangsu Province (Grant No.16KJB520030), the Six Talent Peaks Project of Jiangsu Province (Grant No. 2015-XXRJ-013), and the Priority Academic Program Development of Jiangsu Higher Education Institutions(PAPD).